\documentclass[twocolumn,showpacs,amsmath,amssymb,pra,superscriptaddress,floatfix]{revtex4}
\usepackage{amssymb}
\usepackage{graphicx}
\usepackage{dcolumn}
\usepackage{bm}
\usepackage{psfrag}
\usepackage{slashbox}

\begin{document}

\title[Two-dimensional Rydberg lattice gas]{Creation of collective many-body states and single photons from two-dimensional Rydberg lattice gases}

\author{T. Laycock}
\address{Midlands Ultracold Atom Research Centre - MUARC, The University of Nottingham, School of Physics and Astronomy, Nottingham, United Kingdom}
\author{B. Olmos}
\email{beatriz.olmos-sanchez@nottingham.ac.uk}
\address{Midlands Ultracold Atom Research Centre - MUARC, The University of Nottingham, School of Physics and Astronomy, Nottingham, United Kingdom}
\author{I. Lesanovsky}
\email{igor.lesanovsky@nottingham.ac.uk}
\address{Midlands Ultracold Atom Research Centre - MUARC, The University of Nottingham, School of Physics and Astronomy, Nottingham, United Kingdom}

\date{\today}
\begin{abstract}
The creation of collective many-body quantum states from a two-dimensional lattice gas of atoms is studied. Our approach relies on the van-der-Waals interaction that is present between alkali metal atoms when laser excited to high-lying Rydberg s-states. We focus on a regime in which the laser driving is strong compared to the interaction between Rydberg atoms. Here energetically low-lying many-particle states can be calculated approximately from a quadratic Hamiltonian. The potential usefulness of these states as a resource for the creation of deterministic single-photon sources is illustrated. The properties of these photon states are determined from the interplay between the particular geometry of the lattice and the interatomic spacing.
\end{abstract} \maketitle

\section{Introduction}
The quantum interface between light and an atomic ensemble has attracted much attention during the past decade \cite{Hammerer10}. The reason is rooted in the wide range of possible applications that rely on the coherent coupling between the two systems: Implementation of quantum information processing protocols \cite{Pedersen09,Gorshkov10}, slow-light and electromagnetically induced transparency \cite{Hau99,Fleischhauer00,Schnorrberger09,Unanyan10,Pritchard10} as well as the creation of deterministic and manipulable photon sources \cite{Balic05,Nikoghosyan10,Nielsen10,Pohl10}. In order to create these photon sources, one must be able to create entangled atomic many-body states and map them efficiently into the desired photonic states \cite{Porras08,Scully06,Mazets07,Olmos10-2}. Ultracold atoms represent the ideal tool for creation these atomic states under very clean and well-defined conditions. This is due to the advanced techniques that are nowadays available for the trapping of ultracold atoms and for tailoring the interactions between them \cite{Bloch08}. Highly excited (so-called Rydberg) states \cite{Gallagher94} have proven to be especially useful in this context due to the strong state-dependent interaction between them, allowing the entanglement of atoms separated by several micrometers \cite{Saffman09-2,Mueller09,Weimer10,Wilk10,Isenhower10}.

In a recent work a laser-driven gas of Rydberg atoms trapped on a one-dimensional ring lattice has been proposed as a resource to create collective excitations \cite{Olmos09-3,Olmos10-1}. It was shown that these excitations could be converted into few photons following a mapping scheme proposed in Refs. \cite{Lehmberg70,Porras08,Scully06,Mazets07}. As a result non-classical states of light such as single-photon sources and entangled pairs of photons could be created \cite{Olmos10-2} whose properties where determined by the interplay of the ring geometry and the nature of the delocalized atomic excitations. In this work, we extend this study to a gas of atoms trapped in two-dimensional medium-scale square and triangular lattices. We study thoroughly the properties of collective many-body states in these two lattice types and outline methods of selective excitation. We find that one can create single-photon sources in which the photon is emitted in a superposition of a few well-focused beams. The direction of these beams is mainly determined by the geometry and spacing of the lattice.

The paper is structured as follows. In Section \ref{sec:system} we introduce the system we have in mind and present a derivation of the Hamiltonian that governs its dynamics. In Section \ref{sec:diagonal} we discuss the diagonalization of this Hamiltonian and explain the approximations considered to do so. The symmetry properties of the many-particle eigenstates are analyzed in Section \ref{sec:excitation}, together with a proposal of how to excite experimentally these states. We explain how to map these atomic collective excitations into non-classical states of light and discuss their properties in Section \ref{sec:photon}. Finally, in Section \ref{sec:perturbations} we discuss the validity of the approximations made and take into account disorder caused by quantum uncertainty and/or temperature. We conclude with a summary and outlook in Section \ref{sec:conclusion}.

\section{The system and its Hamiltonian}\label{sec:system}
\begin{figure}[h]
\centering
  \includegraphics[width=\columnwidth]{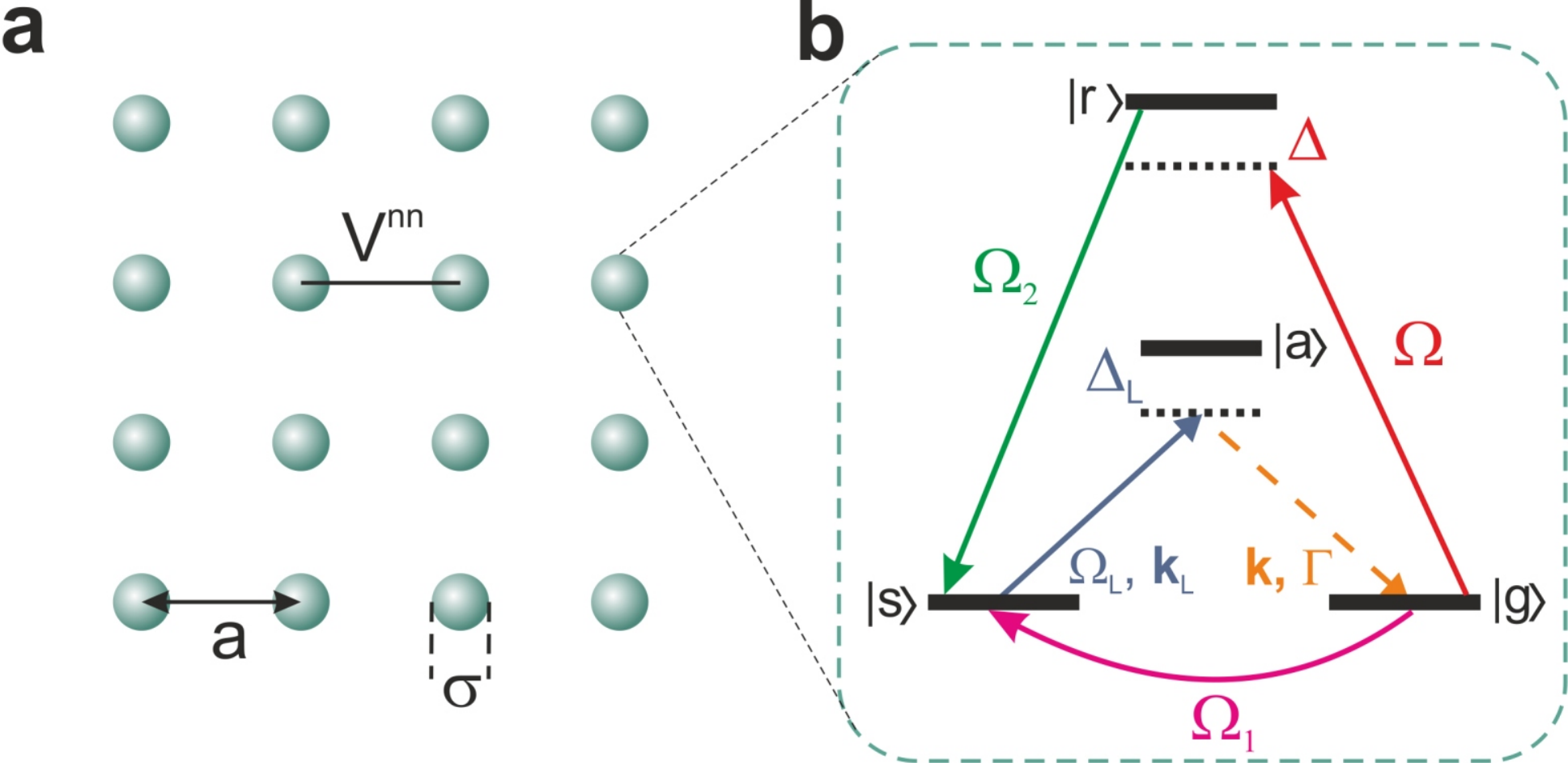}
  \caption{\textbf{a}: Two-dimensional lattice in which the atoms are trapped. The width of the external wave function of each atom is given by $\sigma$ which is considered to be much smaller than the lattice spacing $a$. The largest interaction energy in the system is the one between nearest neighbors $V^\mathrm{nn}$. \textbf{b}: Internal level structure of each of the $N$ atoms on the lattice. The two ground states $\left|g\right>$ and $\left|s\right>$ are coupled to the Rydberg state $\left|r\right>$ by means of two classical lasers with Rabi frequencies $\Omega$ and $\Omega_2$, respectively. A third laser $\Omega_1$ drives the transition $\left|g\right>\rightarrow\left|s\right>$. Finally, $\left|g\right>$, $\left|s\right>$ and an auxiliary state $\left|a\right>$ form a lambda scheme where $\left|s\right>$ is coupled off-resonantly to $\left|a\right>$ and photons are emitted in the transition $\left|a\right>\rightarrow\left|g\right>$.}
  \label{fig:lattice_and_levels}
\end{figure}

In the setup we have in mind a gas of atoms is trapped in a regular two-dimensional lattice with $N$ sites and a lattice spacing $a$ of the order of a few micrometers \cite{Bergamini04,Nelson07,Whitlock09,Kruse10} (see Fig. \ref{fig:lattice_and_levels}a). We consider the limit of a deep lattice where the vibrational states within each site are well-approximated by the eigenstates of a harmonic oscillator and tunneling between the sites is absent. The lattice shall be uniformly occupied with one atom per site (Mott-insulator state) and we assume zero temperature such that each atom populates the ground state of its potential well whose spatial width $\sigma$ is much smaller than the interparticle separation $a$.

The internal (electronic) degree of freedom of each atom is approximated by a two-level system. The ground state $\left|g\right>$ is coupled to a Rydberg ns-state $\left|r\right>$ by means of a laser with Rabi frequency $\Omega$ and detuning $\Delta$ as shown in Fig. \ref{fig:lattice_and_levels}b. For the moment, we do not focus on the two auxiliary levels $\left|s\right>$ and $\left|a\right>$ also shown in this figure. They will be used for the excitation of the many-body states and the mapping of these states into light, that will be explained thoroughly in Sections \ref{sec:excitation} and \ref{sec:photon}, respectively.
Within the two-level approximation we can identify each atom as a spin-$1/2$ degree of freedom where $\left|g\right>_k\equiv\left|\downarrow\right>_k$ and $\left|r\right>_k\equiv\left|\uparrow\right>_k$ \cite{Weimer08,Raitzsch08,Sun08,Olmos09,Lesanovsky10-2}. We can thus employ the Pauli matrices ($\sigma_x^{(k)}$, $\sigma_y^{(k)}$ and $\sigma_z^{(k)}$) to formulate the Hamiltonian governing the excitation dynamics of our system. Using this the Hamiltonian describing the laser excitation process reads in the rotating wave approximation
\begin{equation*}
  H_\mathrm{L}=\sum_{k=1}^N\left[\Omega\sigma_x^{(k)}+\Delta n_k\right],
\end{equation*}
where $n_k=1/2\left(1+\sigma_z^{(k)}\right)$ is the Rydberg number operator of the $k$-th site.

When the atoms are in the Rydberg state, they interact strongly via van-der-Waals interaction $V_\mathrm{vdW}(\mathbf{r})=C_6\times\left|\mathbf{r}\right|^{-6}$, where $\mathbf{r}$ is the separation between the atoms and $C_6$ is the van-der-Waals coefficient \cite{Marinescu95,Singer05}. The interaction Hamiltonian thus becomes
\begin{equation*}
  H_\mathrm{int}=\sum_{k\neq m}V_{km}n_kn_m,
\end{equation*}
with the coefficients
\begin{equation*}
  V_{km}=\frac{C_6}{\left|\mathbf{r}_k-\mathbf{r}_m\right|^6}
\end{equation*}
where $\mathbf{r}_k$ denotes the position of the atom trapped in the $k$-th site. Adding the interaction with the laser, the complete Hamiltonian of the system reads
\begin{equation}\label{eqn:working_hamiltonian}
  H=\sum_{k=1}^N\left[\Omega\sigma_x^{(k)}+\Delta n_k+\sum_{m\neq k}V_{km}n_kn_m\right].
\end{equation}

We assume the external dynamics of the atoms to be frozen. This assumption is well justified because the internal dynamics take place on a much shorter time scale (typically of the order of a hundred nanoseconds versus milliseconds).
We will also assume for the moment that the atoms are infinitely localized at the center of the traps, i.e. $\sigma/a\rightarrow 0$. This very idealized situation can only be achieved approximately in experiment. We will investigate the effect of a finite width of the wave packets in Section \ref{sec:perturbations}.

\section{Diagonalization}\label{sec:diagonal}
In this section, we will show how to obtain the eigenvalues and eigenstates of the Hamiltonian (\ref{eqn:working_hamiltonian}) and discuss the approximations used to obtain them. Throughout this work, we will focus on the regime where the laser driving $\Omega$ is the largest energy scale of the system. In addition, the detuning of the laser $\left|\Delta\right|$ is taken to be much smaller than both the Rabi frequency (laser driving) and the interaction strength such that $\Omega\gg V^\mathrm{nn}\gg\left|\Delta\right|$.

\subsection{Holstein-Primakoff transformation}\label{sec:Hols-Prim}
The Hamiltonian (\ref{eqn:working_hamiltonian}) contains spin operators that obey anti-commutation and commutation relations when they belong to the same and different sites, respectively. Therefore, the underlying algebra is neither bosonic nor fermionic. In order to be able to solve the system at least approximately the spin operators will be expressed in terms of operators that obey a purely bosonic algebra. This is achieved by the Holstein-Primakoff transformation \cite{Holstein40}.

We will make use of a slightly unconventional formulation of this transformation. As $\Omega\gg V^\mathrm{nn}\gg\left|\Delta\right|$, the first term of Hamiltonian (\ref{eqn:working_hamiltonian}) is the dominant one and it is thus convenient to make it diagonal. For this reason, instead of the usual form of the Holstein-Primakoff transformation, we will use a rotated version of it such that $\sigma_z^{(k)}\rightarrow\sigma_x^{(k)}$. This is achieved by means of the unitary operator $U_k=\exp \left[i \frac{\pi}{4} \sigma_y^{(k)} \right]$, after which the Holstein-Primakoff transformation becomes
\begin{eqnarray}\nonumber
  &U_k^\dag\sigma_-^{(k)}U_k=-\frac{1}{2}\left(\sigma_z^{(k)}+i\sigma_y^{(k)}\right)=\sqrt{1-a^\dag_k a_k}\,\,a_k&\\\label{eqn:Holstein-Primakoff}
  &U_k^\dag\sigma_+^{(k)}U_k=-\frac{1}{2}\left(\sigma_z^{(k)}-i\sigma_y^{(k)}\right)=a^\dag_k\sqrt{1-a^\dag_k a_k}&\\\nonumber
  &U_k^\dag\sigma_z^{(k)}U_k=\sigma_x^{(k)}=2a^\dag_k a_k-1,&
\end{eqnarray}
where $a_k$ ($a_k^\dag$) are bosonic operators that create (annihilate) non-interacting bosonic excitations within the system at the $k$-th site.
Considering that the eigenstates of the single-atom operator $\sigma_x^{(k)}$ are
\begin{equation*}
   \left|\pm\right>_k=\frac{1}{\sqrt{2}} \left[\left|g\right>_k\pm\left|r\right>_k\right],
\end{equation*}
the equivalent bosonic states are $\left|-\right>_k\equiv\left|0\right>_k$ and $\left|+\right>_k\equiv\left|1\right>_k=a_k^\dag\left|0\right>_k$, with zero and one bosonic excitation on site $k$, respectively.

By making the substitution (\ref{eqn:Holstein-Primakoff}), the Hamiltonian (\ref{eqn:working_hamiltonian}) takes on a rather complicated form that contains square roots of operators, which are difficult to treat in practical calculations. In order to overcome this difficulty, we make an approximation: our study will be focused on many-body states that carry only a few excitations, i.e. whose total number of bosonic excitations $N_\mathrm{b}=\sum_k\left<a^\dag_ka_k\right>$ is much smaller than the number of sites $N$. In the subspace spanned by these states - denoted by $\langle\!\!\!\langle...\rangle\!\!\!\rangle$ - one has $\langle\!\!\!\langle a_k^\dag a_k \rangle\!\!\!\rangle\ll 1$ for all $k$. Dropping the notation $\langle\!\!\!\langle...\rangle\!\!\!\rangle$ in the following, one can make a Taylor expansion of the square root operator such that, after the normal ordering of the operators, one obtains
\begin{equation*}
  \sqrt{1-a_k^\dag a_k}\approx1-a^\dag_ka_k+\dots
\end{equation*}
We now substitute the operators of eqs. (\ref{eqn:Holstein-Primakoff}) into Hamiltonian (\ref{eqn:working_hamiltonian}) and consistently with the condition $\langle\!\!\!\langle a_k^\dag a_k \rangle\!\!\!\rangle\ll 1$ keep only the first two terms of the expansion of the square root. Furthermore, we neglect all terms of the Hamiltonian that are of higher order than quadratic in $a_k$. The Hamiltonian then assumes a quadratic form
\begin{eqnarray}\label{eqn:Hamiltonian_HP}
H'&\approx&E_0+2\Omega\sum_{k=1}^N a_k^\dag a_k\\\nonumber &-&\frac{\Delta}{2}\sum_{k=1}^N\left(a^\dag_k+a_k\right)-\frac{1}{2}\sum_{k\neq m}V_{km}\left(a^\dag_k+a_k\right)\\\nonumber
&+&\frac{1}{4} \sum_{k\neq m}V_{km}\left(a_k^\dag a^\dag_m+a_k a_m+ a_k^\dag a_m+a^\dag_ma_k\right),
\end{eqnarray}
where the energy offset $E_0$ is given by
\begin{equation}\label{eqn:ground}
  E_0=-N\left(\Omega-\frac{\Delta}{2}\right)+\frac{1}{4}\sum_{m\neq k}V_{km}.
\end{equation}

\subsection{Constrained dynamics}
A closer inspection of the Hamiltonian (\ref{eqn:Hamiltonian_HP}) reveals that it is composed of three distinct parts ($H'=E_0+H_0+H_1+H_2$) which effectuate qualitatively different couplings between states with a fixed number of bosonic excitations. The first part is formed by those terms which, applied to a given state, do not change its number of excitations, i.e.
\begin{equation}\label{eqn:Ham_zero}
  H_0=2\Omega\sum_{k=1}^Na^\dag_k a_k+\frac{1}{4}\sum_{k\neq m}V_{km}\left(a_k^\dag a_m+a^\dag_ma_k\right),
\end{equation}
as they contain the same number of creation and annihilation operators.
Second, there are a number of terms that create or annihilate a single bosonic excitation,
\begin{equation}\label{eqn:Ham_one}
  H_1=-\frac{\Delta}{2}\sum_{k=1}^N\left(a^\dag_k+a_k\right) -\frac{1}{2}\sum_{k\neq m}V_{km}\left(a_k^\dag+a_k\right).
\end{equation}
Finally, there is a third class of terms, which change the total number of bosonic excitations in the system by two:
\begin{equation}\label{eqn:Ham_two}
  H_2=\frac{1}{4}\sum_{k\neq m}V_{km}\left(a_k^\dag a^\dag_m+a_ka_m\right).
\end{equation}

\begin{figure}[h]
\centering
  \includegraphics[width=6cm]{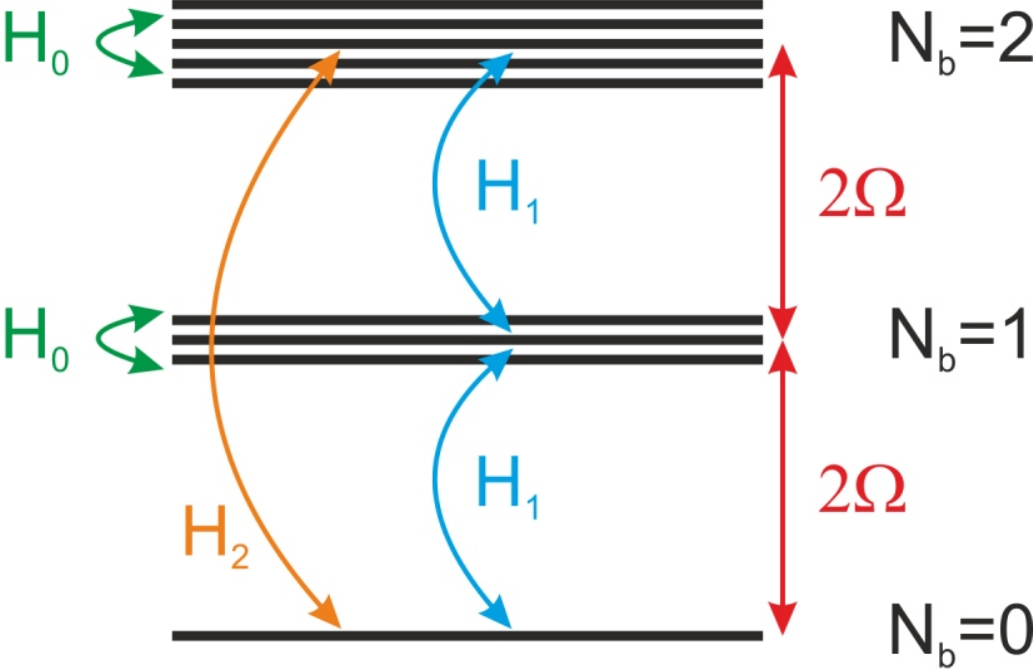}
  \caption{Scheme of the low-energy sector of the spectrum, formed by quasi-degenerate manifolds of states with the same number $N_\mathrm{b}$ of bosonic excitations separated by an energy gap of approximately $2\Omega$. The arrows represent schematically the transitions driven by the different parts of the Hamiltonian (\ref{eqn:Hamiltonian_HP}).}
  \label{fig:spectrum}
\end{figure}
Due to the strong laser driving ($\Omega\gg V^\mathrm{nn}\gg\left|\Delta\right|$), the first term of the Hamiltonian $H_0$ which is proportional to $\Omega$ is clearly dominant and thus determines the coarse structure of the spectrum of the system. This coarse structure is formed by manifolds of quasi-degenerate states separated by energy gaps of approximately $2\Omega$. As it is shown schematically in Fig. \ref{fig:spectrum}, each manifold is formed by states that contain the same total number $N_\mathrm{b}$ of bosonic excitations. Taking this into account, let us now analyze the rest of the terms of the Hamiltonian one by one. The second term in $H_0$ does not change the total number of excitations of the state. Its action produces couplings between states inside the same energy manifold, whose strength is proportional to $V^{\mathrm{nn}}$. The Hamiltonians $H_1$ and $H_2$ create or destroy one and two excitations every time, respectively. Since the states involved in these transitions are separated energetically by approximately $2\Omega$ and $4\Omega$, respectively, they are strongly suppressed. In particular, the corresponding transition rates can be estimated by second order perturbation theory to be proportional to $\Delta^2/\Omega$ and ${V^{\mathrm{nn}}}^2/\Omega$, which in our parameter regime ($\Omega\gg\left|\Delta\right|,V^\mathrm{nn}$) are very small quantities. As a consequence it can be affirmed that, to a very good degree of approximation, the Hamiltonian (\ref{eqn:Ham_zero}) that drives the intra-manifold dynamics also governs the dynamics of the \emph{entire system}. The validity of this approximation will be assessed in Section \ref{sec:perturbations}.

\subsection{Eigenexcitations}
We will now diagonalize the Hamiltonian $H_0$ [eq. (\ref{eqn:Ham_zero})]. Since it is quadratic this is achieved by means of the diagonalization of the matrix $V$, that essentially contains the interaction energies between each pair of atoms. This is achieved by a unitary transformation which reads
\begin{equation*}
  V_{km}=\sum_{ij}M_{ki}D_{i}\delta_{ij}M_{jm}^\dag,
\end{equation*}
where $M_{ki}$ represents the $k$-th entry of the $i$-th eigenvector of $V$ and $D_i$ is its corresponding eigenvalue. Substituting the last expression into (\ref{eqn:Ham_zero}), the diagonalized Hamiltonian becomes
\begin{equation*}
  H_0=\sum_{i=1}^N\epsilon_ib^\dag_i b_i,
\end{equation*}
with the operators
\begin{equation*}
  b^\dag_i=\sum_{k=1}^N M_{ki} a^\dag_k
\end{equation*}
and the eigenvalues $\epsilon_i=2\Omega+\frac{D_i}{2}$ for $i=1,\dots,N$. Hence, the \emph{collective excitations} of the system are formed by a superposition of all the bosonic excitations $a^\dag_k$, with the coefficients of the superposition given by the eigenvectors of $V$.

We can now write down the eigenstates and corresponding energies of the system. The ground state is
\begin{equation}\label{eqn:ground_state}
  \left|0\right>=\prod_{k=1}^N\left|0\right>_k,
\end{equation}
and its energy is $E_0$, given by eq. (\ref{eqn:ground}). The first excited manifold is spanned by the $N$ states that possess one collective bosonic excitation, i.e.
\begin{equation}\label{eqn:first_excited}
  \left|1_i\right>=b^\dag_i\left|0\right>,\qquad E_{1_i}=E_0+\epsilon_i.
\end{equation}
The manifold with $N_\mathrm{b}=2$ is formed by all combinations of two of the $N$ eigenexcitations
\begin{equation}\label{eqn:second_excited}
  \left|2_{ij}\right>=\frac{1}{\sqrt{1+\delta_{ij}}}\,b^\dag_ib^\dag_j\left|0\right>,\qquad E_{2_{ij}}=E_0+\epsilon_i+\epsilon_j,
\end{equation}
with $j\geq i=1\dots N$, so that there exist a total number $N(N+1)/2$ of these doubly excited states.
Following the same procedure, one could think of constructing states with higher numbers of excitations by applying more creation operators $b^\dag$ to the vacuum. However, one has to keep in mind that the Hamiltonian (\ref{eqn:Ham_zero}) gives a valid description of the system only if $\langle\!\!\!\langle a_k^\dag a_k \rangle\!\!\!\rangle\ll 1$, i.e. in the low-occupation sector of the spectrum where only a few bosonic excitations compared to the total number of atoms are present.

\section{Many-particle states: selective excitation and properties}\label{sec:excitation}
Our aim is to selectively excite the many-particle states described in the previous section starting from the experimentally relevant initial state in which all atoms are prepared in the electronic ground state
\begin{equation}\label{eqn:initial}
  \left|\mathrm{init}\right>=\prod_{k}\left|g\right>_k.
\end{equation}
To this purpose, we use the excitation scheme proposed in Refs. \cite{Olmos09-3,Olmos10-1}, that consists of two steps:

First, we want to achieve the ground state of the Hamiltonian $H_0$, $\left|0\right>=\prod_{k=1}^N\left|0\right>_k$, which contains zero bosonic excitations. To do so, we express the single-atom bosonic states in terms of the atomic ones, so that the ground state yields $\left|0\right>=\prod_{k=1}^N\left|-\right>_k$.
We also introduce an additional internal single atom stable state $\left|s\right>_k$ (see Fig. \ref{fig:lattice_and_levels}b). In practice $\left|s\right>_k$ could correspond to one of the states of the hyperfine ground state manifold which is different from $\left|g\right>_k$. Using this state we can prepare the desired state $\left|0\right>$ by means of the two subsequent laser pulses which are depicted in Fig. \ref{fig:lattice_and_levels}b. We choose the first pulse, with Rabi frequency $\Omega_1$, to be resonant with the single-atom transition $\left|g\right>_k\rightarrow\left|s\right>_k$. After applying a $\pi/2$-pulse during a time $\tau_1=\pi/(2\Omega_1)$, this laser is turned off and another one with Rabi frequency $\Omega_2$ and resonant with $\left|s\right>_k\rightarrow\left|r\right>_k$  is shone on the system during a time $\tau_2=\pi/\Omega_2$. This amounts to the sequence
\begin{eqnarray}\label{eqn:mapping}
 \left|\mathrm{init}\right>&=&\prod_k \left|g\right>_k\,\stackrel{\tau_1}{\rightarrow}\,\prod_k \frac{1}{\sqrt{2}}\left[\left|g\right>_k+i\left|s\right>_k\right]\\\nonumber
 &\stackrel{\tau_2}{\rightarrow}& \prod_k \frac{1}{\sqrt{2}}\left[\left|g\right>_k-\left|r\right>_k\right]=\prod_k \left|-\right>_k=\left|0\right>
\end{eqnarray}
which results in the desired preparation of $\left|0\right>$. Here we would like to remark that the strength of the second laser, given by its Rabi frequency $\Omega_2$, has to be large enough to overcome the interaction energy between atoms in the Rydberg state. The required strength will depend strongly on the particular geometry of the system. For example, in a square lattice each atom has four nearest neighbors, so that in this case the condition to be accomplished is roughly $\Omega_2\gg 4V^\mathrm{nn}$, while in a triangular lattice it would be $\Omega_2\gg6V^\mathrm{nn}$ as there are six nearest neighbors for each site.

Once the ground state of the Hamiltonian (\ref{eqn:Ham_zero}) is prepared, the aim is to excite the previously described many-particle states (\ref{eqn:first_excited}) and (\ref{eqn:second_excited}) that contain one and two bosonic excitations, respectively, thus belonging to higher energy manifolds.
The idea proposed in Refs. \cite{Olmos09-3,Olmos10-1} is to induce transitions between states with $\Delta N_\mathrm{b}=\pm1$ making use of the term of the Hamiltonian proportional to the detuning, i.e. the first term of (\ref{eqn:Ham_one}),
\begin{equation}\label{eqn:Ham_detuning}
  H_\Delta=-\frac{\Delta}{2}\sum_{k=1}^N\left(a_k^\dag+a_k\right),
\end{equation}
which creates or annihilates one bosonic excitation at a time. The problem is that, as we discussed before, these transitions are highly suppressed due to the large value of the energy gap - approximately $2\Omega$, see Fig. \ref{fig:spectrum} - between the manifolds. To overcome this difficulty one can introduce a time-dependent detuning of the form
\begin{equation*}
  \Delta(t)=\Delta_0\cos{\left(\omega_\Delta t\right)},
\end{equation*}
which will therefore act as a coherent oscillating source that couples the two many-body states considered when the oscillation frequency $\omega_\Delta$ matches the corresponding energy gap between them. The intensity of the transition between the two states considered is calculated in each case by working out the square of the transition matrix elements $I=\left|\left<i\right|H_\Delta\left|f\right>\right|^2$, where $\left|i\right>$ and $\left|f\right>$ represent the initial and final state, respectively.

Let us first consider the excitation from the ground state $\left|0\right>$ to the first excited manifold of states (\ref{eqn:first_excited}) that contain a single collective bosonic excitation.
The intensity yields in this case
\begin{equation}\label{eqn:I1}
  I_1(i)\equiv\left|\left<0\right|H_\Delta\left|1_i\right>\right|^2= \frac{\left|\Delta_0\right|^2}{16}\left|\sum_{k=1}^N M_{ki}\right|^2,
\end{equation}
where the Rotating Wave Approximation (RWA) has been made and $\epsilon_i=2\Omega+\frac{D_i}{2}$ is the corresponding energy gap between the state $\left|1_i\right>$ and the ground state. Note that the RWA implies that the value of the energy gap - which is roughly proportional to $\Omega$ - is much larger than the detuning, i.e. $\Delta_0\ll\Omega$.

Following the same procedure, one can calculate the intensity of the transition between the one- and two-excitation states, that yield
\begin{eqnarray}\label{eqn:I2}
  I_2(i;jk)&\equiv&\left|\left<1_i\right|H_\Delta\left|2_{jk}\right>\right|^2\\\nonumber
  &=&\frac{\delta_{ik}I_1(j)+\delta_{ij}I_1(k)+ \delta_{ij}\delta_{ik}\sqrt{I_1(j)I_1(k)}}{1+\delta_{jk}}.
\end{eqnarray}
Note that this expression already tells us that the intensity of the transition between the two states will be zero unless one of the excitations in the two-boson state is the same as the single one originally present, i.e. unless $i=j$ or $i=k$. Other selection rules that stem from the particular geometry of the system will be analyzed in the following section.

\subsection{Selection rules}\label{sec:selection}
\begin{figure}[h]
\centering
  \includegraphics[width=\columnwidth]{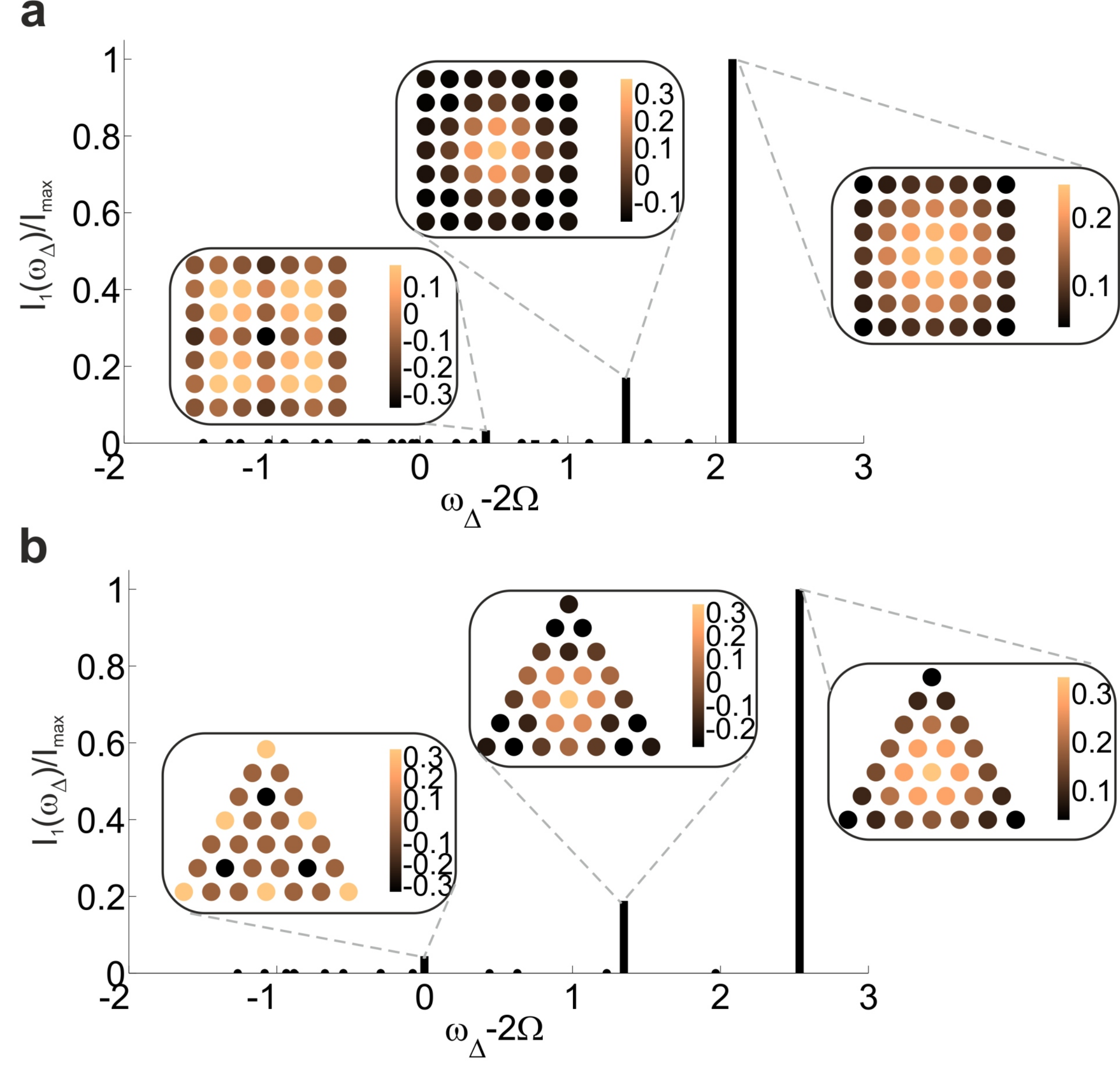}
  \caption{Normalized absorption profile for the transition $\left|0\right>\rightarrow\left|1_i\right>$. Only the states that belong to the totally symmetric subspace $A_1$ have a non-zero probability of being excited. The three states associated to the highest intensities are shown \textbf{a}: in a $7\times7$ square lattice and \textbf{b}: in a equilateral triangular lattice with $7$ sites per side. The energies are given in units of $V^\mathrm{nn}$.}
  \label{fig:I1}
\end{figure}

\begin{figure}[h]
\centering
  \includegraphics[width=\columnwidth]{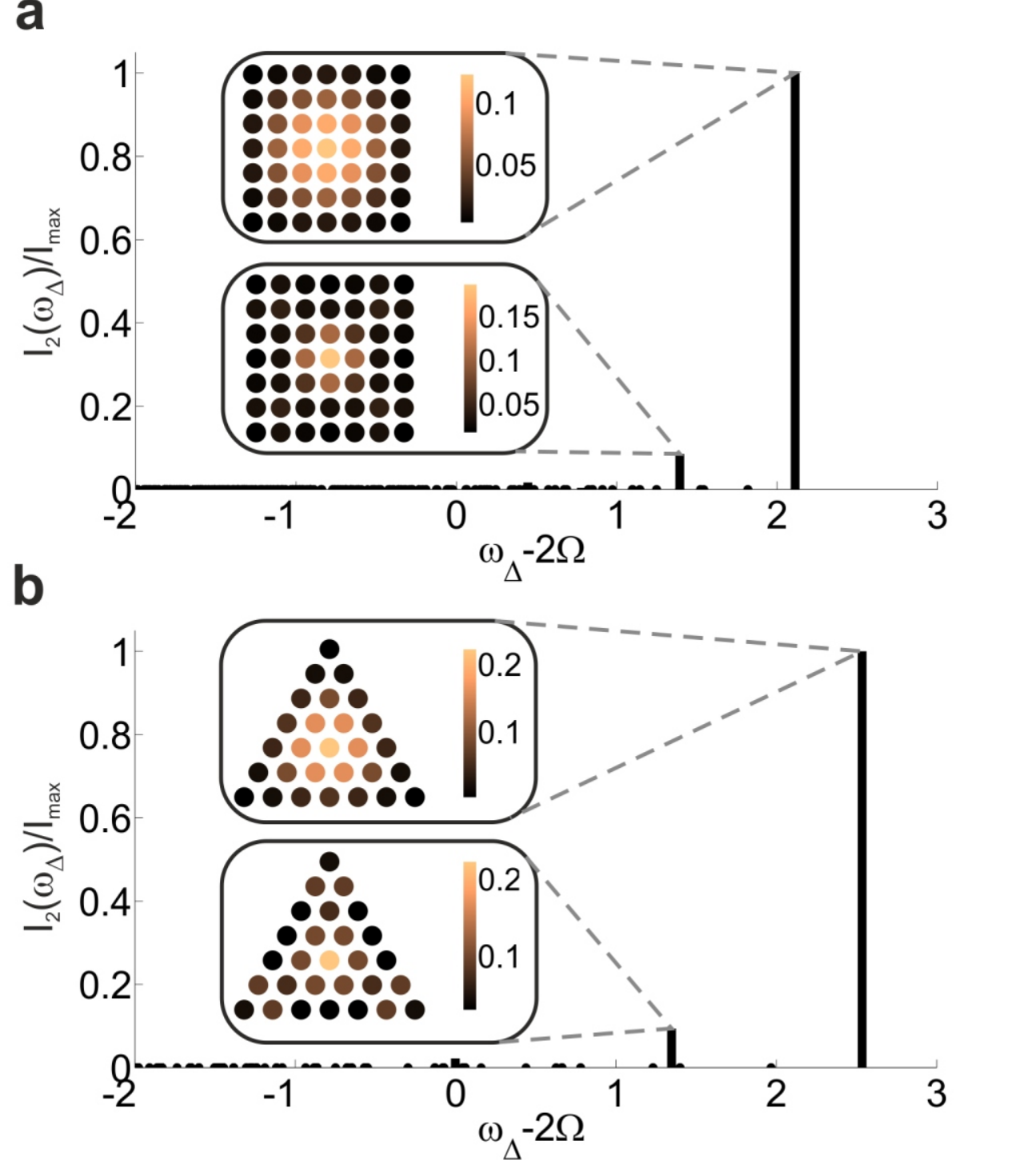}
  \caption{Normalized absorption profile for the transition $\left|1_N\right>\rightarrow\left|2_{ij}\right>$. Only the states that belong to the totally symmetric subspace $A_1$ have a non-zero probability of being excited. The density of bosons of the two states associated to the highest intensities are shown \textbf{a}: in a $7\times7$ square lattice and \textbf{b}: in an equilateral triangular lattice with $7$ sites per side. The energies are given in units of $V^\mathrm{nn}$.}
  \label{fig:I2}
\end{figure}

When calculating the intensity of the transitions between the different manifolds following Eqs. (\ref{eqn:I1}) and (\ref{eqn:I2}), we observe that only for a few states they have a non-zero value (see Figs. \ref{fig:I1} and \ref{fig:I2}). The reason for this can be found considering the particular symmetries of the system. Let us consider here two examples of geometries: a square and an equilateral triangular lattice.
\begin{figure}[h]
\centering
  \includegraphics[width=\columnwidth]{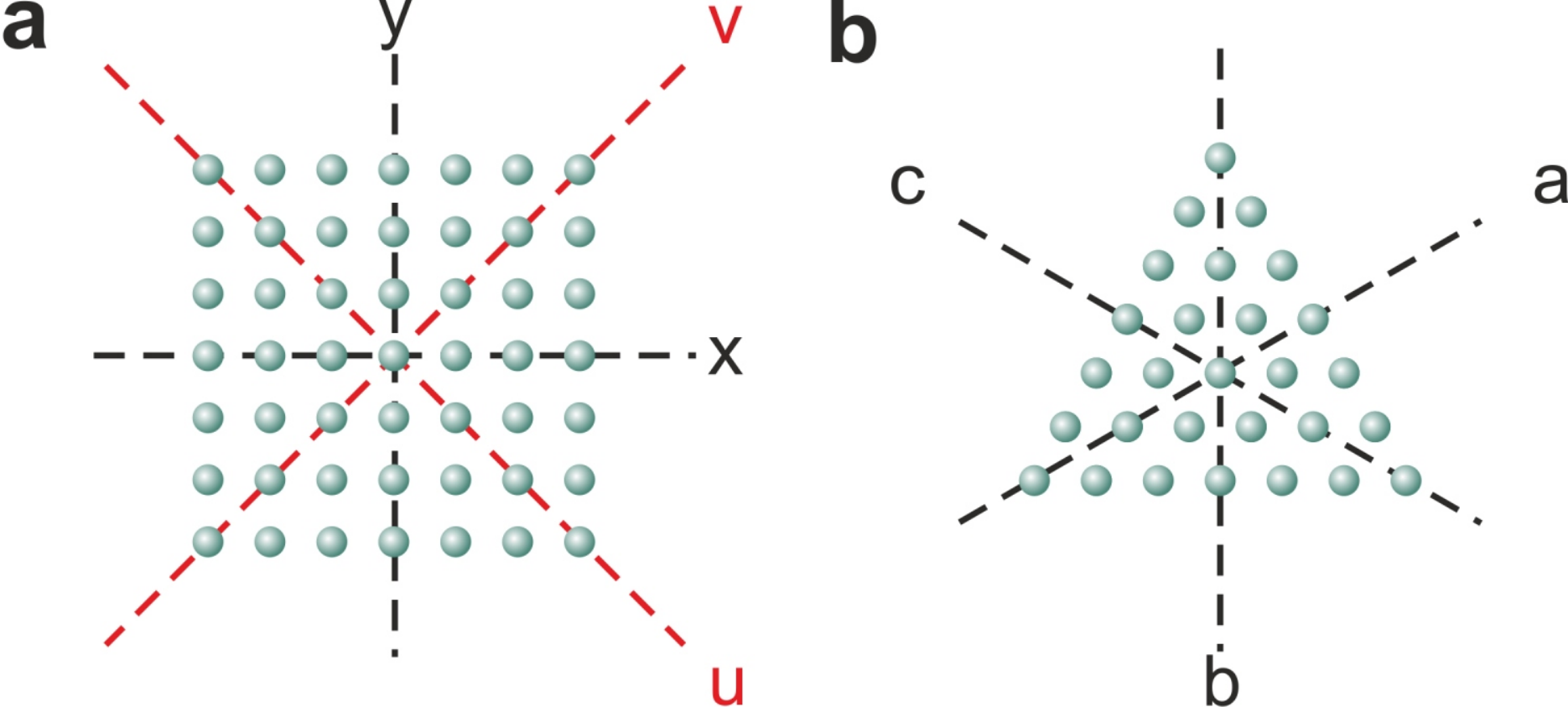}
  \caption{Scheme of two lattices and their symmetries. \textbf{a}: The group $C_{4v}$ that contains the symmetries of a square is formed by 8 elements: Rotations of the lattice by $0,\pi/2,\pi$ and $3\pi/2$; flip about the $x$ and $y$ axes; flip about the diagonals $u$ and $v$. \textbf{b}: The group $C_{3v}$ contains the symmetries of an equilateral triangle, formed by 6 elements: Rotations of $0,2\pi/3$ and $4\pi/3$ and flip about the axes $a,b$ and $c$ that go through the vertices of the triangle perpendicular to the opposite side.}
  \label{fig:symmetries}
\end{figure}

The symmetry groups corresponding to the square and the equilateral triangle are $C_{4v}$ and $C_{3v}$, respectively. These groups are formed by the following transformations (see Fig. \ref{fig:symmetries}a and b):
\paragraph{Square}
\begin{itemize}
  \item $C^4_n$, rotations of $2\pi n/4$, with $n=0,1,2,3$
  \item $F_x,F_y$, flip about the vertical and horizontal axes
  \item $F_u,F_v$, flip about the two main diagonals
\end{itemize}
\paragraph{Equilateral triangle}
\begin{itemize}
  \item $C^3_n$, rotations of $2\pi n/3$, with $n=0,1,2$
  \item $F_a,F_b,F_c$, flip about the three axes trough the vertices
\end{itemize}
One can observe that both the complete Hamiltonian $H$ (\ref{eqn:working_hamiltonian}) and $H_\Delta$ (\ref{eqn:Ham_detuning}) conserve all the symmetries corresponding to the system's geometry. As a consequence, if the initial state is an eigenstate with respect to the previous symmetry operators, the time evolution will be restricted to the subspace spanned by the states with the same quantum number with respect to the cited operators.
Actually, both the experimentally relevant initial state $\left|\mathrm{init}\right>$ and the ground state $\left|0\right>$ where no bosonic excitation is present, are eigenstates of the operators that represent \emph{all} the rotations and flip operations that form the corresponding groups with eigenvalue $+1$. This is true for both the square and the triangular lattice.
The states that have eigenvalue $+1$ with respect to all the operators representing the group elements belong to the subspace $A_1$, called \emph{totally symmetric}. Thus, only the states that belong to this totally symmetric subspace can be accessed in the course of the time evolution under Hamiltonians $H$ and $H_\Delta$.

Keeping these selection rules in mind, let us now consider the excitation of the states with a single collective bosonic excitation from the ground state $\left|0\right>$ and calculate the intensity of these transitions. An example is given in Fig. \ref{fig:I1}a and b for a square lattice of $N=49$ sites and a triangular one with $N=28$, respectively.
Here we observe that, as a consequence of the selection rules, the intensity is always zero except for a very restricted set of final states. These states belong to the subspace $A_1$, i.e. they are symmetric (eigenvalue $+1$) with respect to all the symmetry operations corresponding to the particular geometry of the lattice.
Finally, we also observe that the state with highest eigenenergy is the one with the largest intensity. This is observed for all sizes of the system (determined by $N$) that have been studied and for both geometries considered. This state reads $\left|1_N\right>=\sum_k M_{kN}a_k^\dag\left|0\right>$, with $M_{kN}\geq 0$, and is the closest state to a uniformly shared collective excitation, i.e. a spin wave without spatially varying phase factors.

Let us now discuss briefly the excitation of the states that carry two bosonic excitations. In particular, let us assume that we have already excited the state $\left|1_N\right>$ which corresponds to the largest intensity $I_1(N)$ and calculate the transitions from that state to the doubly-excited manifold. As we already obtained in Eq. (\ref{eqn:I2}), the intensity is zero unless one of the two excitations corresponds to the initial one, e.g. in this case $b^\dag_N$. This corresponds to only $N$ possible states out of the total $N(N+1)/2$ available. Moreover, the selection rules will reduce the number of accessible states even more due to the fact that only the states that belong to $A_1$ (totally symmetric states) can be accessed from $\left|1_N\right>$, which also belongs to $A_1$.
In particular, a state that carries two excitations only belongs to the symmetry subset $A_1$ when the two excitations that form it belong to the same symmetry subspace. As a consequence, the number of doubly-excited states with non-zero transition intensity is the same as the number of singly excited ones with $I_1\neq 0$, which can also be directly inferred from Eq. (\ref{eqn:I2}).
To understand this better, let us study an example. For a $3\times3$ ($N=9$) square lattice there are only three eigenexcitations that belong to $A_1$, which are $b^\dag_1,b^\dag_5$ and $b^\dag_9$. Let us consider that we are interested in knowing which states can be then accessed from the state $\left|1_9\right>=b^\dag_9\left|0\right>$. In principle any combination of the eigenexcitations $b^\dag_jb^\dag_k$ with $j\geq k=1\dots 9$ should be accessible. However, we know from the result (\ref{eqn:I2}) that only if $j=9$ or $k=9$ the intensity of the transition is non-zero. Finally, the selection rules impose that only the combinations of the three eigenexcitations that belong to $A_1$, i.e. $\left|2_{19}\right>$, $\left|2_{59}\right>$ and $\left|2_{99}\right>$ can be accessed from $\left|1_9\right>$ with $I_2\neq0$.
In Fig. \ref{fig:I2}a and b we show the corresponding intensities for the same two systems considered before, i.e. $7\times7$ square lattice and equilateral triangle with $7$ sites on each side, respectively. In this case, in the insets of the figures show the bosonic density in each site $n$, i.e. $\left<2_{jk}\right|a^\dag_n a_n\left|2_{jk}\right>$, for the two states corresponding to the highest intensity.

\subsection{Mapping to a stable state}

Once the many-particle atomic states have been excited, we need to map them into a stable configuration. This is necessary due to the limited lifetime of the Rydberg states and if one wants to manipulate the corresponding state further. To this end we make use of the stable hyperfine states introduced in the previous section, $\left|s\right>_k$ (see Fig. \ref{fig:lattice_and_levels}b).
Let us remind that each of the single-atom bosonic states $\left|1\right>_k,\left|0\right>_k$ is equivalent to the atomic single-atom one $\left|+\right>_k$ and $\left|-\right>_k$, respectively (see Section \ref{sec:Hols-Prim}). As a consequence, by inverting the sequence of laser pulses given in (\ref{eqn:mapping}), we achieve the mapping
\begin{equation*}
  \left|0\right>_k\rightarrow\left|g\right>_k\qquad \left|1\right>_k\rightarrow\left|s\right>_k,
\end{equation*}
so that the excitations are stored now in the two hyperfine ground states. Expressed in terms of the atomic states, the many-body states studied and given by (\ref{eqn:first_excited}) and (\ref{eqn:second_excited}) become
\begin{equation}\label{eqn:first_atomic}
  \left|\Psi^{(1)}_{i}\right>=\sum_{k=1}^N M_{ki}\sigma_{sg}^{(k)}\left|\mathrm{init}\right>
\end{equation}
and
\begin{equation}\label{eqn:second_atomic}
  \left|\Psi^{(2)}_{ij}\right>=\sum_{k,l=1}^N M_{ki}M_{lj}\sigma_{sg}^{(k)}\sigma_{sg}^{(l)}\left|\mathrm{init}\right>,
\end{equation}
respectively, with $\sigma_{sg}^{(k)}=\left|s\right>_k\left<g\right|$.

\section{Application: Single photon sources}\label{sec:photon}
So far, we found the low-lying eigenexcitations in the lattice and described the properties of the many-body states that can be accessed experimentally by controlling the laser parameters. In this section, we will show how to map the collective atomic excitations into quantum states of light \cite{Porras08,Scully06,Mazets07}.

\subsection{General aspects of the atom-photon mapping}
The mapping from the atomic excitations encoded in the two hyperfine ground states $\left|g\right>_k$ and $\left|s\right>_k$ to non-classical states of light has been thoroughly investigated in \cite{Porras08,Olmos10-2}. In the following we will briefly outline the basic idea.

We consider the level scheme shown in the lower part of Fig. \ref{fig:lattice_and_levels}b where the two ground states $\left|g\right>_k$ and $\left|s\right>_k$ together with an auxiliary state $\left|a\right>_k$ form a lambda scheme. The storage state $\left|s\right>_k$ is coupled off-resonantly to the state $\left|a\right>_k$ via a classical laser field with Rabi frequency and detuning $\Omega_\mathrm{L}$ and $\Delta_\mathrm{L}$ respectively. Eventually, a photon is emitted to the radiation field by means of the transition $\left|a\right>_k\rightarrow\left|g\right>_k$. Note that we do not consider the decay from the state $\left|a\right>_k$ back to $\left|s\right>_k$, a condition that can be accomplished by an appropriate choice of the atomic levels \cite{Porras08}.

This setup allows us to establish a direct mapping between any given atomic state - encoded in $\left|g\right>_k$ and $\left|s\right>_k$ - and its corresponding photonic one after a time much longer than the lifetime $\tau=\Gamma^{-1}$ of the intermediate state $\left|a\right>_k$ with $\Gamma$ being the corresponding decay rate. In particular, one can show that a single atomic excitation is mapped to a single photon state as
\begin{equation*}
  \sigma_{sg}^{(k)}\,\stackrel{t\gg\tau}{\rightarrow}\, \sum_{\mathbf{q}\nu}g_{k\mathbf{q}\nu}(t)a^\dag_{\mathbf{q}\nu}
\end{equation*}
where $a^\dag_{\mathbf{q}\nu}$ creates a photon with momentum $\mathbf{q}$ and polarization $\nu$. The details of the coupling are contained in the function $g_{k\mathbf{q}\nu}(t)$, that - in the limit of $t\gg\tau$ - is given by
\begin{equation}\label{eqn:g}
  g_{k\mathbf{q}\nu}(t)=-iK_{\mathbf{q}\nu}e^{-i\left(\omega t-\mathbf{k}_\mathrm{L}\cdot\mathbf{r}_k\right)} \sum_{n,\gamma=1}^{N}e^{-i\mathbf{q}\cdot\mathbf{r}_\gamma} \frac{\chi_{\gamma n}\chi_{nk}^{-1}}{i\omega-\kappa_n}.
\end{equation}
Let us clarify the meaning of each term appearing in the previous expression. First, $\mathbf{k}_\mathrm{L}$ is the momentum of the laser and $\omega=\left|\mathbf{q}\right|/c$ is the energy of the emitted photon, which also appears in the coefficient
\begin{equation*}
  K_{\mathbf{q}\nu}=\frac{\Omega_\mathrm{L}}{\Delta_\mathrm{L}} \sqrt{\frac{\omega}{2\epsilon_0{\cal V}}}\,\mathbf{d_{\mathrm{ga}}}\cdot\mathbf{e}_{\mathbf{q}\nu}.
\end{equation*}
Here, ${\cal V}$ is the quantization volume, $\epsilon_0$ the vacuum permitivity, $\mathbf{d_{\mathrm{ga}}}$ the dipole operator of the $\left|g\right>_k\rightarrow\left|a\right>_k$ transition and $\mathbf{e}_{\mathbf{q}\nu}$ the polarization vector ($\mathbf{q}\cdot\mathbf{e}_{\mathbf{q}\nu}=0$).
Finally, in eq. (\ref{eqn:g}) $\boldsymbol{\chi}_n$ and $\kappa_n$ represent respectively the eigenvectors and eigenvalues of the operator that governs the atomic dynamics in the process of light emission from an ensemble of atoms. This operator accounts for light scattering at multiple atoms as well as  interatomic dipole-dipole interactions and depends mainly on the relative orientation of the atomic transition dipole moments and the ratio between the interparticle separation and the wavelength of the laser, $a/\lambda_\mathrm{L}$ \cite{Lehmberg70}.

\subsection{Creation of single photons}
We will now use this mapping to obtain the single-photon states that are created when using the atomic states (\ref{eqn:first_atomic}) that contain one collective excitation as a resource. The produced single-photon states are
\begin{equation*}
  \left|\Phi^{(1)}_{i}\right>=\sum_{k=1}^N \sum_{\mathbf{q}\nu}M_{ki}g_{k\mathbf{q}\nu}(t)a^\dag_{\mathbf{q}\nu}\left|\mathrm{vac}\right>,
\end{equation*}
where $\left|\mathrm{vac}\right>$ represents the photon vacuum. In order to characterize the properties of the photonic states, we will study the angular photonic distribution, i.e.
\begin{equation*}
  {\cal I}(\theta,\phi)=\frac{{\cal V}}{(2\pi c)^3}\int_0^\infty \sum_{\nu}\left<n_{\mathbf{q}\nu}\right>\omega^2d\omega,
\end{equation*}
where $n_{\mathbf{q}\nu}=a^\dag_{\mathbf{q}\nu} a_{\mathbf{q}\nu}$ is the number of photons with momentum $\mathbf{q}$ and polarization $\nu$. This quantity provides us with the average photon number in the direction of emission parameterized by $\theta$ and $\phi$.

Let us now specify the parameters of our setup. We consider the lattice to be lying in the $xy$-plane. The dipoles of the transitions $\left|g\right>_k\rightarrow\left|a\right>_k$ are all aligned and oriented perpendicular to the lattice, i.e. $\hat{\mathbf{d}}_{ga}\parallel\hat{\mathbf{z}}$. This can be ensured by an appropriate choice of the atomic levels. Finally, we consider the situation where the incident direction of the laser momentum $\mathbf{k}_\mathrm{L}$ is also perpendicular to the lattice plane, i.e. $\hat{\mathbf{k}}_\mathrm{L}\parallel\hat{\mathbf{z}}$. The only degree of freedom left is - apart from the specific geometry of the lattice - the ratio between the interparticle separation given by the lattice spacing $a$ and the wavelength of the laser $\lambda_\mathrm{L}$. We will see in the following how the change of this ratio modifies dramatically the properties of the emitted photon.

\begin{figure}[h]
\centering
  \includegraphics[width=\columnwidth]{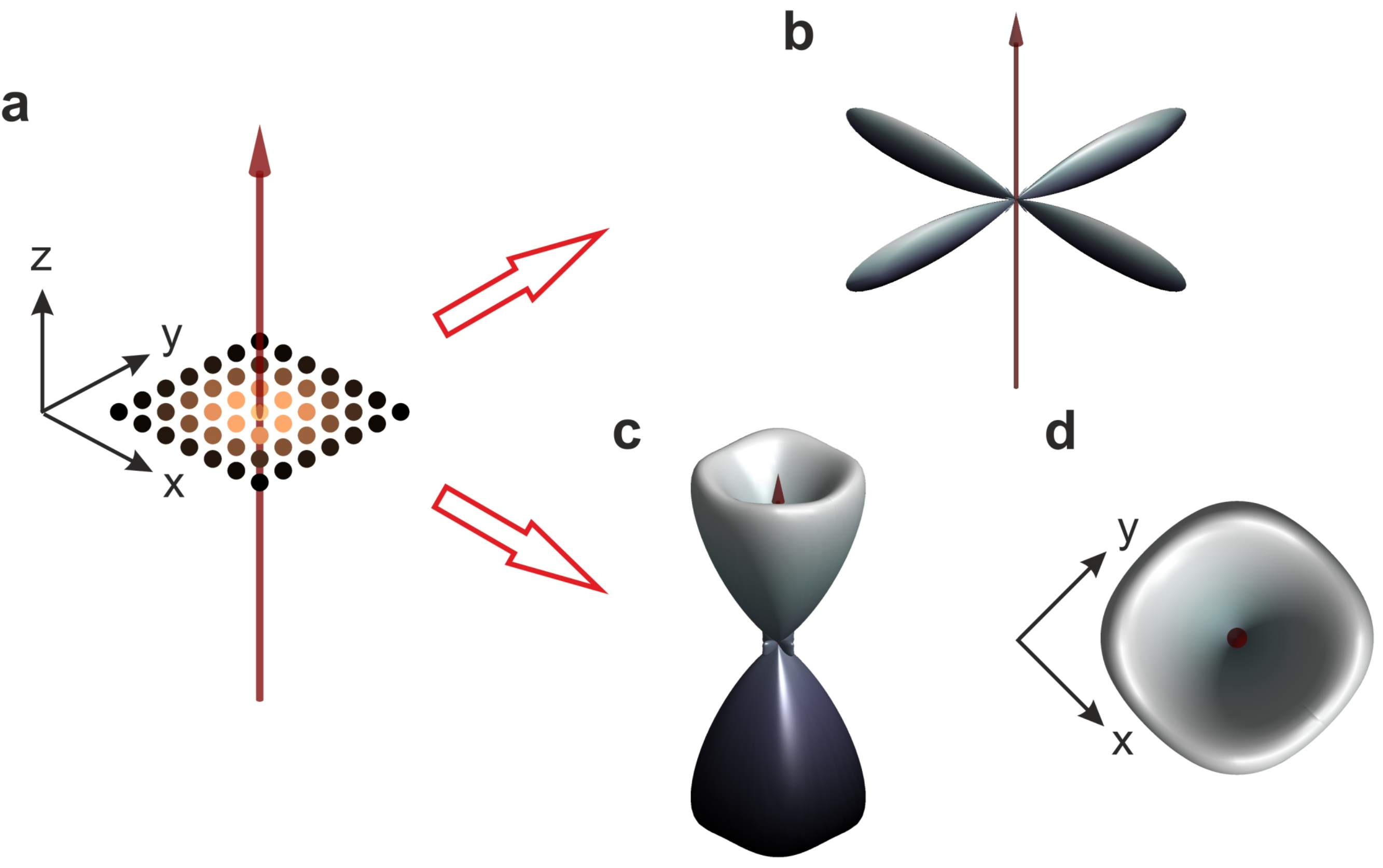}
  \caption{\textbf{a}: The atomic state $\left|\Psi_{N}^{(1)}\right>$ in the case of a $7\times7$ lattice is mapped into a single photon state by means of a laser whose direction is given by the red arrow. \textbf{b}: When $a/\lambda_\mathrm{L}=0.9$, the photon is emitted in a superposition of four beams that are perpendicular to the sides of the square lattice. \textbf{c}: The angular distribution is strikingly different in the case $a/\lambda_\mathrm{L}=0.25$, where the emission occurs into two pyramidal-shaped beams with square base, as it can be observed in \textbf{d}.}
  \label{fig:light_square}
\end{figure}
\begin{figure}[h]
\centering
  \includegraphics[width=\columnwidth]{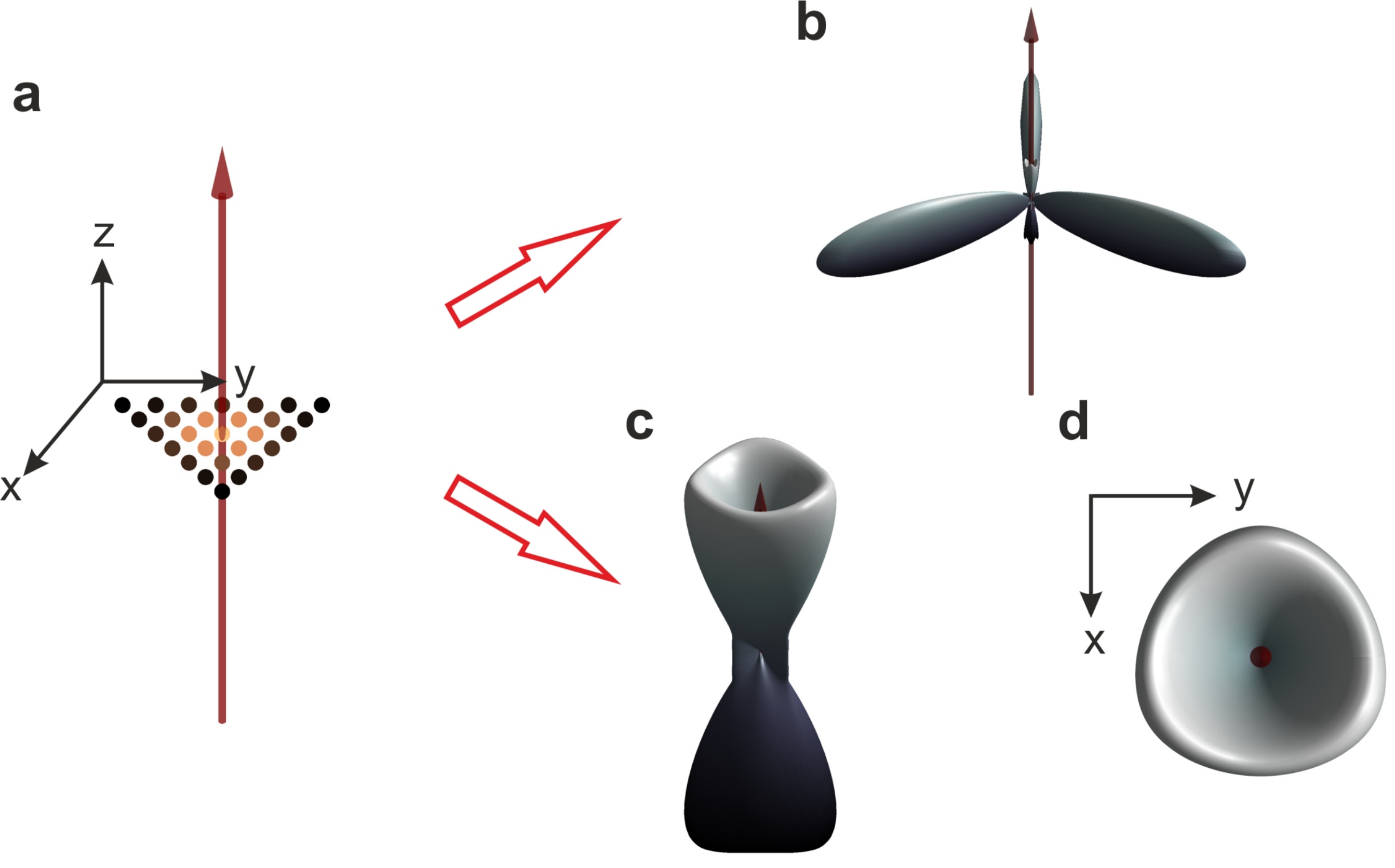}
  \caption{\textbf{a}: The atomic state $\left|\Psi_{N}^{(1)}\right>$ in the case of a triangular lattice of side $7$ is mapped into a single photon state by means of a laser whose direction is given by the red arrow. \textbf{b}: When $a/\lambda_\mathrm{L}=0.9$, the photon is emitted in a superposition of three beams that are perpendicular to the sides of the triangular lattice. \textbf{c}: The angular distribution is strikingly different in the case $a/\lambda_\mathrm{L}=0.4$, where the emission occurs into two pyramidal-shaped beams with triangular base, as it can be observed in \textbf{d}.}
  \label{fig:light_triangle}
\end{figure}
In Figs. \ref{fig:light_square} and \ref{fig:light_triangle} we show two examples of these single photon sources. In both cases we have chosen to map to a photonic state the atomic many-particle state with a single collective excitation with the highest energy $\left|\Psi_N^{(1)}\right>$, which also correspond to the largest transition intensities for each geometry (see Figs. \ref{fig:I1}a and b). The corresponding states are represented in Figs. \ref{fig:light_square}a and \ref{fig:light_triangle}a for a square and a triangular lattice, respectively. The red arrow represents the direction of the incident mapping laser. For each geometry we show the angular distribution of the emitted photon for two different values of the ratio $a/\lambda_\mathrm{L}$.

In the case of a square lattice and for $a/\lambda_\mathrm{L}=0.9$ (see Fig. \ref{fig:light_square}b) we see that the photon is emitted in a superposition of four directions, each one perpendicular to the sides of the square lattice. For the same ratio and a triangular lattice (Fig. \ref{fig:light_triangle}b), the photon is emitted in a superposition of three directions, again perpendicular to the sides of the lattice. With this example we see clearly how the geometry of the lattice plays an important role in the properties of the photonic state. These states are particularly interesting since the photon in this case is not only emitted in a highly directional form but also perpendicularly with respect to the incident laser, which would avoid noise problems in the photon detection process.

In order to illustrate how the ratio $a/\lambda_\mathrm{L}$ also affects the properties of the state, we show in Fig. \ref{fig:light_square}c the mapping of the same atomic state in the square lattice but this time with $a/\lambda_\mathrm{L}=0.25$. We observe that in this case the photon is emitted in a superposition of directions that form two cone-shaped beams that still conserve the information of the geometry of the lattice as can be seen in Fig. \ref{fig:light_square}d. The same numerical experiment has been done in the case of the triangular lattice in Figs. \ref{fig:light_triangle}c and d for $a/\lambda_\mathrm{L}=0.4$ with a similar result.

\section{Validity of the approximations and robustness of the solutions}\label{sec:perturbations}

So far, we have considered a very idealized situation. In particular, we have focused on a regime where the number of excitations is much smaller than the total number of atoms, the laser driving is much stronger than the interaction between the Rydberg states and the atoms are distributed in a regular lattice. In this section we will test the validity of the first two approximations made in order to diagonalize the Hamiltonian, i.e. the Holstein-Primakoff approximation and the assumption that the dynamics is constrained to subspaces with equal excitation number. In addition, we will check the robustness of our results with respect to uncertainties in the atomic positions. This will bring us closer to the experimental realistic situation with quantum uncertainty and/or finite temperature.

\subsection{Validity of Holstein-Primakoff and constrained dynamics approximations}

Throughout Section \ref{sec:diagonal} where we performed the diagonalization of the Hamiltonian (\ref{eqn:working_hamiltonian}) two approximations were made. First, the Holstein-Primakoff approximation $N_\mathrm{b}\ll N$ allowed us to neglect all terms of the Hamiltonian of higher than quadratic order. Second, we considered that in the regime of strong laser coupling, i.e. $\Omega\gg V^{\mathrm{nn}}$, the dynamics of the entire system is approximately driven by the Hamiltonian (\ref{eqn:Ham_zero}), which conserves the number of excitations.
In order to test the validity of these approximations, we consider now the complete Hamiltonian (\ref{eqn:working_hamiltonian}) and rewrite it as
\begin{equation*}
  H\approx E_0+H_0+H_\mathrm{pert}
\end{equation*}
where $E_0$ is the unperturbed energy of the ground state (\ref{eqn:ground}) and $H_0$ is the unperturbed Hamiltonian (\ref{eqn:Ham_zero}). The perturbation to this Hamiltonian can be divided into three terms
\begin{equation*}
  H_\mathrm{pert}=H_1+H_2+H_\mathrm{HP},
\end{equation*}
where $H_1$ and $H_2$ are given in (\ref{eqn:Ham_one}) and (\ref{eqn:Ham_two}), respectively and the third contribution $H_\mathrm{HP}$ contains the terms of third and fourth order in $a_k$ that were neglected by the Holstein-Primakoff approximation
\begin{eqnarray*}
  H_\mathrm{HP}&=&\frac{1}{2}\sum_{k=1}^N\left\{\left(a_k^\dag a^\dag_ka_k+a^\dag_ka_ka_k\right)\left[\Delta+\sum_{m\neq k}V_{km}\right]\right\}\\
  &-&\frac{1}{2}\sum_{k\neq m}V_{km}\left(a^\dag_ka^\dag_ka^\dag_ma_k+a^\dag_ka_ka_ka_m\right.\\
  &&+\left.a^\dag_ka^\dag_ka_ka_m+a^\dag_ka^\dag_ma_ka_k\right).
\end{eqnarray*}
Let us now calculate the second order energy shifts caused by the three terms of $H_\mathrm{pert}$ to the unperturbed energies of the ground and first excited states of the Hamiltonian $H_0$. These contributions vanish only in the limits $\left|\Delta\right|/\Omega\rightarrow0$, $V^\mathrm{nn}/\Omega\rightarrow0$ and $N_\mathrm{b}/N\rightarrow0$. Here we consider that $\Delta=0$, and we calculate the contributions for a finite value of the other two ratios.

We focus first on the corrections to the energy of the ground state. $H_\mathrm{pert}$ only contains terms that couple states that differ by $\Delta N_\mathrm{b}=0,\pm1,\pm2$. As a consequence, only the states of the first and second excited manifold (\ref{eqn:first_excited}) and (\ref{eqn:second_excited}) contribute to the second order shift to the energy of the ground state, given by
\begin{equation*}
  E_0^{(2)}=\sum_{i=1}^N\frac{\left|\left<1_i\right|H_1\left|0\right>\right|^2}{E_0-E_{1_i}} +\sum_{i\leq j}\frac{\left|\left<2_{ij}\right|H_2\left|0\right>\right|^2}{E_0-E_{2_{ij}}}.
\end{equation*}
Note that, since $H_\mathrm{HP}\left|0\right>=0$ only $H_1$ and $H_2$ contribute to the correction of the energy $E_0^{(2)}$.
This correction yields
\begin{equation}\label{eqn:correction_ground}
  E_0^{(2)}=-\frac{1}{8}\sum_{i=1}^N\frac{\left|D_i\right|^2\left(1+4\left|\sum_k M_{ki}\right|^2\right)}{4\Omega+D_i}
\end{equation}
and depends only on the values of $\Omega$, $N$ and the eigenvectors and eigenvalues of the interaction matrix $V$ given by $M_{ki}$ and $D_i$, respectively. The relative error $E_0^{(2)}/E_0\times100$ introduced by these shifts for a square lattice of side $L\equiv\sqrt{N}$ and several values of the ratio $\Omega/V^\mathrm{nn}$ are given in Table \ref{tab:ground}. One can observe that the dependence with $\Omega/V^\mathrm{nn}$ can be very well approximated to be proportional to $\left(\Omega/V^\mathrm{nn}\right)^{-2}$, as it was already considered in Section \ref{sec:diagonal}. Note that for $\Omega/V^\mathrm{nn}=20$ the shift of the energy of the ground state is below $0.6\%$ for all sizes of the lattice considered, so that we can consider the unperturbed energy a very good approximation to the exact solution.
The dependence with the size of the lattice has a more complicated form, although one can observe clearly that the perturbation becomes stronger for larger sizes $N$. This is expected since the number of states that are coupled to the ground state also increases, and so do the number of terms of the sum (\ref{eqn:correction_ground}).
\begin{table}
\centering
\begin{tabular}{|c||c|c|c|c|c|c|c|}
\hline
\backslashbox{$\Omega/V^\mathrm{nn}$}{$L$} & 3 & 4 & 5 & 6 & 7 & 8 & 9 \\
\hline
\hline
5 & 5.0 & 6.5 & 7.5 & 8.2 & 8.7 & 9.1 & 9.4 \\
10 & 1.2 & 1.6 & 1.8 & 2.0 & 2.1 & 2.2 & 2.3 \\
20 & 0.31 & 0.40 & 0.45 & 0.49 & 0.52 & 0.54 & 0.56 \\
50 & 0.049 & 0.063 & 0.072 & 0.079 & 0.083 & 0.087 & 0.090 \\
\hline
\end{tabular}
\caption{Error (in $\%$) introduced by the second order energy shifts to the ground state energy $E_0$ given in (\ref{eqn:ground}) for $\Delta=0$, different values of $\Omega/V^\mathrm{nn}$ and sides of the square lattice given by $L\equiv\sqrt{N}$.}\label{tab:ground}
\end{table}

We follow the same procedure to calculate the second order energy shifts of the energies of the manifold with one bosonic excitation. In this case, we also have to consider the coupling of the $\left|1_\alpha\right>$ states to the triply-excited ones by $H_2$ and $H_\mathrm{HP}$:
\begin{eqnarray*}
  E_{1_\alpha}^{(2)}&=&\sum_{\alpha=1}^N\frac{\left|\left<0\right|H_1\left|1_\alpha\right>\right|^2}{E_{1_\alpha}-E_0} +\sum_{i\leq j}\frac{\left|\left<2_{ij}\right|H_1+H_\mathrm{HP}\left|1_\alpha\right>\right|^2}{E_{1_\alpha}-E_{2_{ij}}}\\
  &&+\sum_{i\leq j\leq k}\frac{\left|\left<3_{ijk}\right|H_2+H_\mathrm{HP}\left|1_\alpha\right>\right|^2}{E_{1_\alpha}-E_{3_{ijk}}}.
\end{eqnarray*}
The final expression of the energy shift is in this case much more unwieldy and thus we will not include it here. As an example, we choose again the state of the single excitation manifold with highest energy, i.e. $\left|1_N\right>$, that corresponds to the largest intensity of the transition from the ground state (the results are almost identical when choosing other states). The percentages of variation of the energy of this state due to the second order shift are shown in Table \ref{tab:first} for several values of $\Omega/V^\mathrm{nn}$ and $L$ and $\Delta=0$. Note that the values are similar to the ones of the ground state. The dependence on the laser strength follows again the expected $\left(V^\mathrm{nn}/\Omega\right)^{-2}$ behavior.
The correction to the energy due to the term $H_\mathrm{HP}$ decreases with increasing number of sites $N$. This is expected, since the contribution of the terms of higher than quadratic order vanishes when $N/N_\mathrm{b}\rightarrow0$.
On the other hand, the energy shifts due to the other two terms of the perturbation Hamiltonian $H_1$ and $H_2$ become larger when increasing the number of states (the size of the system). The interplay of these two opposing tendencies makes the dependence of the quantity $E_{1_N}^{(2)}/E_{1_N}\times100$ less pronounced than in the case of the ground state.
\begin{table}
\centering
\begin{tabular}{|c||c|c|c|c|c|c|c|}
\hline
\backslashbox{$\Omega/V^\mathrm{nn}$}{$L$} & 3 & 4 & 5 & 6 & 7 & 8 & 9 \\
\hline
\hline
5 & 7.2 & 8.1 & 8.7 & 9.1 & 9.4 & 9.7 & 9.9 \\
10 & 1.6 & 1.8 & 2.0 & 2.1 & 2.2 & 2.3 & 2.3 \\
20 & 0.37 & 0.44 & 0.49 & 0.52 & 0.54 & 0.56 & 0.58 \\
50 & 0.057 & 0.069 & 0.077 & 0.082 & 0.086 & 0.089 & 0.092 \\
\hline
\end{tabular}
\caption{Error (in $\%$) introduced by the second order energy shifts to the singly-excited state $\left|1_N\right>$ energy $E_{1_N}$ given in (\ref{eqn:first_excited}) for $\Delta=0$, different values of $\Omega/V^\mathrm{nn}$ and sides of the square lattice given by $L\equiv\sqrt{N}$.}\label{tab:first}
\end{table}

\subsection{Uncertainty in the atomic positions, $\sigma/a\neq 0$}

We have considered so far that the atoms populate the harmonic oscillator ground state and that the atoms were infinitely localized, so that the width of the external wave function was negligible compared to the interparticle separation, i.e. $\sigma/a\rightarrow 0$.
Let us now consider a small but finite value of this ratio given by the inevitable uncertainty of the atomic position due to the finite strength of the confinement and finite temperature of the system.

\begin{figure}[h]
\centering
  \includegraphics[width=\columnwidth]{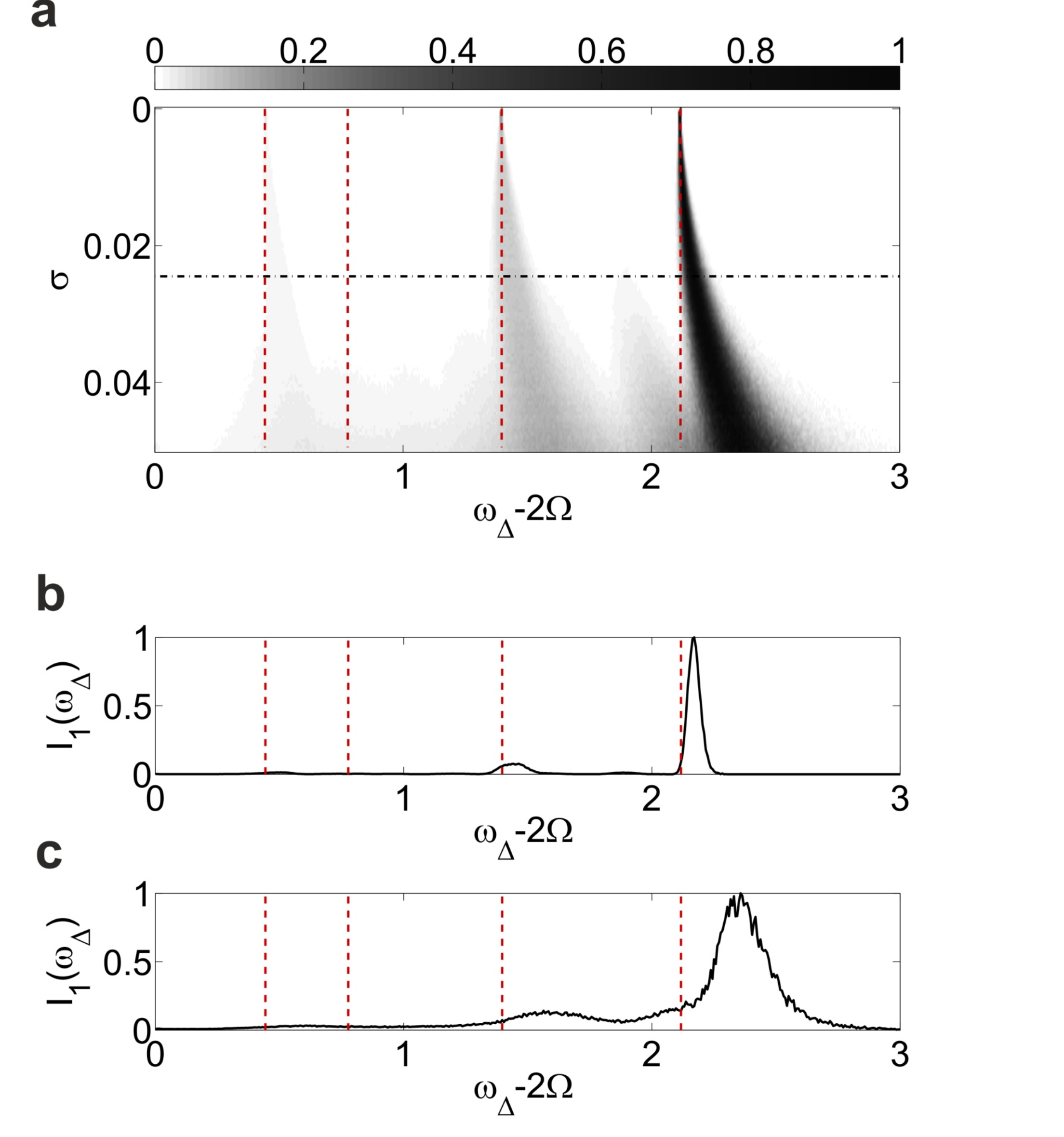}
  \caption{\textbf{a}: Normalized absorption profile for the transition from the ground to the first excited states when the positions of the atoms in the lattice are randomly distributed around the center of each site with standard deviation $\sigma$ (in units of $a$). The red vertical lines show the transition lines for $\sigma=0$. The horizontal black line indicates the cut shown in \textbf{b}: for $\sigma=0.025$. \textbf{c}: The same cut for $\sigma=0.05$. The results are averaged over 10000 realizations.}
  \label{fig:sigma}
\end{figure}

We are particularly interested in how the intensities of the transitions from the ground state to the single excitation manifold $I_1$ given by Eq. (\ref{eqn:I1}) change when considering this uncertainty. To do so, we distribute randomly the position of the atoms around the center of each site of the lattice with a finite standard deviation $\sigma\neq 0$. As a consequence, the symmetry of the system is lifted, the selection rules discussed in Section \ref{sec:selection} are not applicable anymore and thus more states of the single excitation manifold become accessible from the ground state. In Fig. \ref{fig:sigma} we show the resulting absorption profile after averaging over 10000 realizations up to a maximum value of $\sigma/a=0.05$ for a square lattice of $7\times 7$ sites. We observe that as we increase the uncertainty in the position of the atoms the initial sharp lines (indicated by the red vertical lines) become broader and blue-shifted. The asymmetry of the shift can be explained as a consequence of the geometry of the lattice. Let us first consider the configuration with $\sigma=0$ and that only one atom in the lattice can move from the center of the site. This atom can move in 6 possible directions, i.e. $\pm \hat{\mathbf{x}}$, $\pm \hat{\mathbf{y}}$ or $\pm \hat{\mathbf{z}}$. If it moves in the $xy$-plane, it will end up closer to at least one of its neighboring atoms. Since the interaction grows as $1/r^{6}$, the energy in this case will be increased. Only the configurations in which the atom moves in the $\pm \hat{\mathbf{z}}$ directions will give rise to a lower interaction energy. In total 4 out of the 6 possibilities give rise to a larger energy, which explains the overall shift to the blue in the profile. When the atom is located in one of the boundary sites, the number of configurations that reduce the energy is 3 out of 6. As the number of sites on the lattice $N$ increases, the relative number of atoms on the boundaries decreases and, as a consequence, the shift to the blue is more pronounced. This phenomenon is consistent with the observations obtained from the numerical simulations.

\section{Conclusions and outlook}\label{sec:conclusion}

In this work, we have studied the excitation properties of a laser-driven gas of highly excited atoms confined to a two-dimensional lattice.
In order to solve the system's Hamiltonian we have made two approximations. First, we have restricted our study to the states with a small number of excitations compared with the total number of atoms in the system. Moreover, we have focused on the regime where the strength of the laser is much larger than the interaction between the atoms in the Rydberg state. After these approximations the resulting Hamiltonian could be solved by diagonalizing a small matrix which allowed us to obtain the many-body eigenstates of the system and their corresponding energies.

Moreover, we have described a protocol of how to experimentally access the many-body states in the low-occupation sector of the spectrum by the use of an oscillating detuning. Here we have found that only the states with the same symmetry properties as the ground state of the system can actually be excited.
We have shown how to map these collective many-body atomic states into non-classical single photon sources and observed that the interplay between the interatomic distance and the wavelength of the emitted photon changes dramatically the angular distribution of the photon state. In particular, we have identified a regime in which the photon is emitted in a superposition of well-separated beams, whose number and relative direction depends strongly on the geometry (square or triangular) of the underlying lattice.
This feature has potential applications in quantum information processing, especially to form quantum networks.

Finally, we have considered a small uncertainty in the positions of the atoms on each site of the lattice due to the finite temperature and the quantum nature of the external atomic wave function, which brings us closer to the actual experimental situation. The intensity of the transition from the ground state to the first excited manifold has been calculated for varying values of the uncertainty in the position and we have found a systematic broadening and blue-shift in the profile.

\section*{Acknowledgments}
The authors acknowledge funding by EPSRC. B. Olmos also acknowledges funding by Fundaci\'on Ram\'on Areces.


\end{document}